\documentstyle[aps,prl]{revtex}
\input epsf.sty
\begin{document}

\twocolumn[\hsize\textwidth\columnwidth\hsize\csname@twocolumnfalse\endcsname

\draft
\title{Random-Mass Dirac Fermions in an Imaginary 
Vector Potential (II): \\
Long-Range Correlated Random Mass}
\author{Koujin Takeda\cite{Takeda}}
\address{Institute for Cosmic Ray Research, University of Tokyo, Kashiwa, 
Chiba, 277-8582 Japan
}
\author{Ikuo Ichinose\cite{Ichinose}}
\address{Department of Electrical and Computer Engineering, 
Nagoya Institute of Technology, Gokiso, Showa-ku, 
Nagoya, 466-8555 Japan
}
%\date{, 2000}
\maketitle
\begin{abstract}
In the previous paper, 
we studied the random-mass Dirac fermion in one
dimension by using the transfer-matrix methods and by introducing an
imaginary vector potential in order to calculate the localization lengths.
Especially we considered effects of the nonlocal but short-range 
correlations of the random mass.
In this paper, we shall study effects of the long-range correlations of 
the random mass especially on the delocalization transition.
The results depend on how randomness is introduced in the Dirac mass.

\end{abstract}

]

%\begin{multicols}{2}
\section{Introduction}
In the previous papers \cite{PPs,KI1}, we studied the random-mass 
Dirac fermions 
in one dimension by using the transfer-matrix methods (TMM)
and imaginary-vector potential method (IVPM).
We calculated the density of states, the typical and mean localization
lengths, and the multifractal scalings
as a function of the energy and correlation length of the
random mass.
The results are in good agreement with the available analytical
calculations \cite{SUSY}.
We also obtained the relation between the correlation length
of the random mass and the typical localization length
for the short-range correlations \cite{KI1}.

In this paper, we shall consider effects of the long-range correlations,
especially on the delocalization transition.
This problem is interesting.
Very recently finite mobility edges are 
observed in the one-dimensional Anderson model with the long-range correlated
potentials though it is widely believed that
(almost) all states are localized in spatial one dimension 
\cite{lyra,izrailev}.
Similar phenomena are observed also in the aperiodic Kronig-Penney
model \cite{IzKrUl}.
In contrast to the above models, the system of
the random-mass Dirac fermions belongs to the universality class of the 
{\em chiral}
orthogonal ensemble and therefore the extended states exist in the 
band center though the others are localized for the white-noise
random mass.
Long-range correlations of the disorder may change this feature.
This problem has not been addressed in detail so far.
As the model is closely related with the random-bond XY model, the random 
Ising model, etc., the results are interesting both theoretically and 
experimentally.

This paper is organized as follows.
The model and numerical methods are explained in the previous paper \cite{KI1}.
In Sec.2, we shall explain how to make a long-range correlated random
Dirac mass numerically.
We shall consider two types of the telegraphic random mass, i.e.,
in the first one magnitudes of the random mass
are long-range correlated random variables
(LRCRV) with fixed interval distances between kinks
whereas in the second one interval distances between kinks are 
LRCRV with a fixed random mass magnitude.
As we show in later, physical results are different in the above two cases.
As far as we know, this possibility has not been found in the previous studies
on the random systems.
In Sec.3, numerical results are given.
We calculate the localization lengths directly by using the TMM and IVPM
though in most of studies of one-dimensional random models
elaborated techniques like the Hamiltonian mapping,
the renormalization group, etc. are used for calculating the Lyapunov exponent.
Before obtaining the localization lengths in the present methods, 
we have various landscapes of the wave functions themselves 
and obtain a physical picture of the ``metal-insulator"
transition from them.
Section 5 is devoted to conclusion.

%%%%%%%%%%%%%%%%%%%%%%%%%%%%%%%%%%%%%%%%%%%%%%%%%%%%%%%%%%%%%%%%%%%%%%%5
\section{Models and Long-range correlated random mass}

Hamiltonian of the random-mass Dirac fermion is given by
\begin{eqnarray}
{\cal H}&=&\int dx \psi^\dagger h\psi,\\
h&=&-i\sigma^z \partial_x +m(x)\sigma^y,
\end{eqnarray}
where $\vec{\sigma}$ are the Pauli matrices and $m(x)$ is the
telegraphic random mass. 
In this paper we shall consider the long-range correlated
configurations of $m(x)$ and study delocalized states by
the TMM and IVPM.
Numerical methods which generate long-range correlated potentials was
recently invented by Izrailev {\it et.al.} and Herbut
\cite{izrailev,herbut}.
We briefly review it for we shall use it for the present studies.

Let us consider a spatial lattice and random potential $\epsilon(n)$
sitting on the sites which has correlation $\chi(n)$,
\begin{equation}
  [\ \epsilon (m) \; \epsilon (n) \ ]_{\rm ens}= C\; \chi{(|m-n|)},
\label{disordercor2}
\end{equation}
where $n$ and $m$ are site indices, and $C$ is a constant.
Here we suppose that $\chi(n)$ is given and define 
$\tilde{c}(k)$ from $\chi(n)$ as follows,
\begin{equation}
  \tilde{c}(k) = \tilde{\chi}(k)^{1/2},
\end{equation}
where $\tilde{\chi}(k)$ is the discrete Fourier transform of $\chi(n)$;
\begin{equation}
\tilde{\chi}(k)= \sum^{\infty}_{n=-\infty} \chi(n)e^{ikn},
\end{equation}
We also introduce another random site potential $r(n)$ 
which is distributed uniformly in the range $[-1,1]$ with
the delta-function white-noise correlation,
\begin{equation} 
 [\ r(m) \; r(n)\ ]_{\rm ens} \propto \delta_{m,n}.
\end{equation}
 Then we can construct random potential $\epsilon(n)$ as follows
 by using the white-noise 
 potential $r(m)$ and the inverse Fourier transform of $\tilde{c}(k)$,
\begin{equation}
 \epsilon (n) = C^{'} \sum^{\infty}_{m=-\infty} r (m)\; c(m-n),
\end{equation} 
where $C^{'}$ is another constant.
However, we cannot perform the infinite summation in the numerical 
calculation, and therefore
we replace it with the finite summation as follows,
\begin{equation}
\label{eq:construct}
 \epsilon (n) = C^{'} \sum^{k/2}_{m=-k/2} r (m)\; c(m-n).
\end{equation} 
 where $k$ is a finite number which is 
 much larger than the number of sites.\footnote{We use the 
 periodic boundary condition for the numerical calculation
 in subsequent sections.}
 We set $k \simeq 10^{4}+\mbox{(number of sites)}$ in our 
 numerical study.
 
 We can easily verify that the correlator (\ref{disordercor2})
 is satisfied from these variables.
 If we choose $c(m)$ as
\begin{equation}
 c (m) = \frac {1}{|m|^{\alpha}},
\end{equation}
then the correlator of the random potential becomes
\begin{equation}
[\ \epsilon (m) \; \epsilon (n) \ ]_{\rm ens}  \sim \frac{1}
 {|m-n|^{2 \alpha -1} }.
 \label{ep-correlation}
\end{equation}
 With this procedure, 
 we can generate power-law correlated random 
 potential.

In the actual calculation,
there is a subtle point in the normalization of the random potential
$\epsilon(n)$ in Eq.(\ref{ep-correlation})
when we vary the system size $L$ \cite{comment}.
We determine the normalization of $\epsilon(m)$ (namely the value of 
 $C^{'}$) in such a way that the typical value of $\epsilon(m)$ does not
 change.
Then as a result, the strength of correlation
 does not change for various system size $L$ if we fix $|m-n|$ in
 Eq.(\ref{ep-correlation}).
 
The above method generates the random potential at each site
(Fig.1).
Therefore, if we fix the
 distances between kinks and vary the value of $m(x)$
for generating random $m(x)$, we can
 directly use this method. 
 We can extend TMM easily 
 to the case of random values of $|m(x)|$.
 Numerical studies of this system will be reported in Sec.3.A.
\begin{figure}
\label{fig:example1}
\begin{center}
\unitlength=1cm

\begin{picture}(15,4)
\centerline{
\epsfysize=4cm
\epsfbox{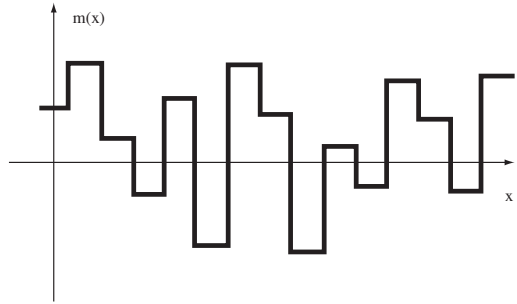}
}
\end{picture}
\vspace{0mm}
\caption{An example of $m(x)$}
\end{center}
\end{figure}
On the other hand, we can also use the above method for generating
random distances between kinks with fixed value of $|m(x)|$.
This is the subject in Sec.3.B.
Physical results are different in the above two cases.

%%%%%%%%%%%%%%%%%%%%%%%%%%%%%%%%%%%%%%%%%%%%%%%%%%%%%%%%%%%%%%%%%%%%%%%
\section{Long-range correlated disorders}
\subsection{Random mass with random magnitude}
We focus on the system with power-law correlated random mass
in this subsection.
As we explained at the end of Sec.2, we first consider
the random mass of random magnitude with fixed distances between
kinks.
Prototype of the configuration is given in Fig.2.
In this case $m(x)$ is generated rather similarly to
$\epsilon(n)$ in Sec.2 though the continuum space is considered here
instead of the spatial lattice.
As in the previous paper \cite{KI1}, we use the 
IVPM and TMM in order to calculate the localization
lengths.

We first calculate the correlation
 $[ m(x) m(0) ]_{\rm ens}$
 numerically which is expected to
exhibit the power-law decay.
The result is shown in Fig.3.
The correlator actually shows the algebraic decay, though the decay
powers are slightly smaller than the expected values 2$\alpha-1$ (denoted
by $\alpha_{pw}$).
We think that this is due to the finiteness of numbers of kinks
and/or {\em finite} Fourier transformation in the scheme which
we explained in the previous section. 

\begin{figure}
\label{fig:example2}
\begin{center}
\unitlength=1cm

\begin{picture}(15,3.5)
\centerline{
\epsfysize=4cm
\epsfbox{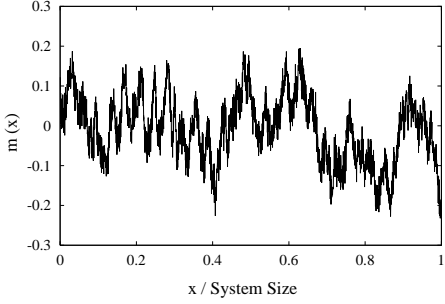}
}
\end{picture}
\vspace{0mm}
\caption{An example of $m(x)$ in the case of power-law correlated
      random mass : 
 We set $\alpha_{pw}=0.5$ and 2000 kinks.}
\end{center}
\end{figure}

\begin{figure}
\label{fig:powerlaw}
\begin{center}
\unitlength=1cm

\begin{picture}(15,4)
\centerline{
\epsfysize=4cm
\epsfbox{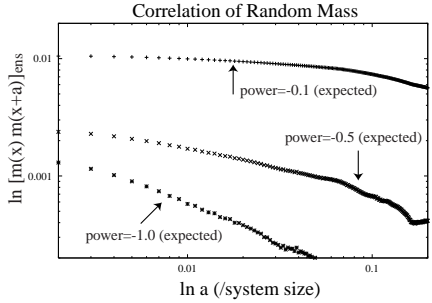}
}
\end{picture}
\vspace{0mm}
\caption{Correlation of $m(x)$ in the case of power-law correlated
      random mass : 
 Points for each $\alpha_{pw}$(power index) are almost on one line.}
\end{center}
\end{figure}

Next we show the calculations of the typical localization length
of the eigenstates.
(We obtain ``typical" localization length by averaging localization lengths
 of the eigenstates within a small range of energy.)
As we vary the magnitude of the random mass $|m(x)|$ in the present
calculation, we should remark the following facts.
If energy $E$ is larger than $|m(x_a)|$, the eigenfunction corresponding
that energy is given by the plain wave near $x \sim x_a$.
On the other hand if $E<|m(x_a)|$, the eigenfunction is a linear
combination of exponentially decaying and increasing functions.
Therefore for eigenstate with $E> \mbox{Max}\;\{|m(x)|\}$, the wave function
is given by a plain wave everywhere and the plain waves are connected
at each position of (anti-)kinks, i.e, jumps of $m(x)$. 
This, of course, does {\em not} mean that all eigenstates with 
$E> \mbox{Max}\;\{|m(x)|\}$ are necessarily extended.

Calculation of the localization lengths {\em vs.} energy
is given for comparatively large $\alpha_{pw}$ in Figs.4 and 5.
There are many sharp peaks as in the case of Gaussian distributed
random mass which was studied in the previous paper \cite{KI1}.
To calculate typical localization length as a function of energy,
we average localization length of all states around each peak
as in the previous cases.
The results in Figs.4 and 5 show
the existence of the mobility edge at a {\em finite
energy} which roughly equals to $\mbox{Max}\;\{|m(x)|\}$.
\begin{figure}
\begin{center}
\unitlength=1cm
\begin{picture}(15,4)
\centerline{
\epsfysize=4cm
\epsfbox{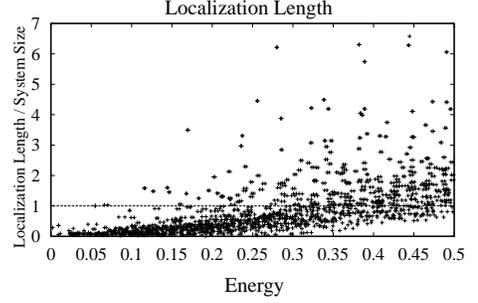}
}
\end{picture}
\begin{picture}(15,4.5)
\centerline{
\epsfysize=4cm
\epsfbox{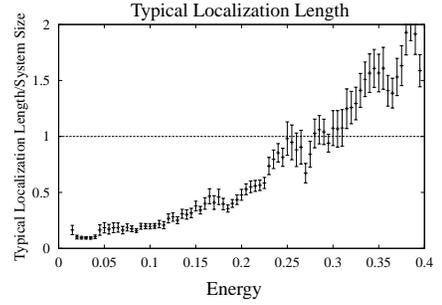}
}
\end{picture}
\end{center}
\vspace{0mm}
\caption{Plots of localization length and typical localization length (1):
 We set $L=300$, $\alpha_{pw} = 0.01$ and $2000$ kinks in the system.
$\mbox{Max}\;\{|m(x)|\}$ is about $0.29$.}
\label{fig:power1}
\end{figure}
\vspace{1cm}
\begin{figure}
\begin{center}
\unitlength=1cm
\begin{picture}(15,3.5)
\centerline{
\epsfysize=4cm
\epsfbox{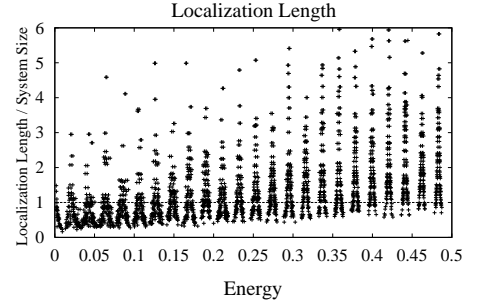}
}
\end{picture}
\begin{picture}(15,4.5)
\centerline{
\epsfysize=4cm
\epsfbox{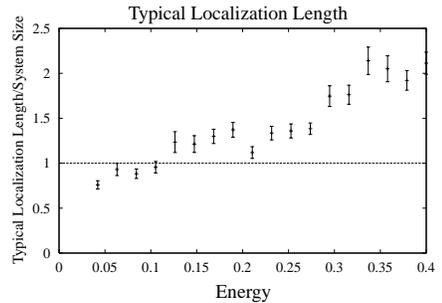}
}
\end{picture}
\end{center}
\vspace{0mm}
\caption{Plots of localization length and typical localization length (2):
 We set $L=300$, $\alpha_{pw}=2$ and $500$ kinks in the system.
 $\mbox{Max}\;\{|m(x)|\}$ is about $0.13$.}
\label{fig:power2}
\end{figure}
They also show that the above result concerning the existence of
the mobility edge holds regardless of the value of $\alpha_{pw}$. 
Typical localization length is an increasing function of energy.
This is in sharp contrast to the previous result for the system of
random kink-distances in which the localization length is a decreasing
function of energy.\footnote{It is interesting to study in detail
the low-energy behaviour of the localization length in the present
system.}

We also studied the system size dependence of typical localization length.
The result is shown in Figs.6 and 7.
We found that energy of the mobility edge becomes larger as we let 
the system size larger and it can exceed the value
 $\mbox{Max}\;\{|m(x)|\}$ in large systems.
($\mbox{Max}\;\{|m(x)|\}$ for different size systems are almost the same.)
 Localized states (states whose localization length is smaller than the system
 size) are also observed in the range $E>$Max$\{ |m(x)| \}$. 

From the above results, we can conclude that
the Maximum value of $|m(x)|$ does not
give the mobility edge or critical energy of the Anderson transition, 
whereas states which have small energy are all localized in
large systems. (We also expect that the states near $E=0$ are always
extended exceptionally.) 
This conclusion does not depend on the value of $\alpha_{pw}$.

\begin{figure}
\begin{center}
\unitlength=1cm
\begin{picture}(15,3.5)
\centerline{
\epsfysize=4cm
\epsfbox{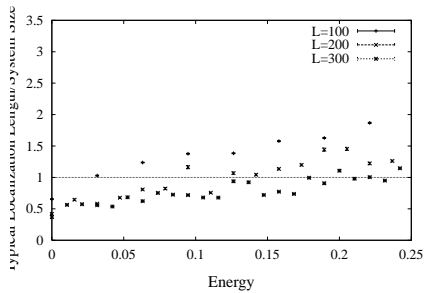}
}
\end{picture}
\end{center}
\vspace{0mm}
\caption{System size dependence of typical localization length (1):
 We set $\alpha_{pw}=1.0$. Distance between kinks is $0.5$ and
 $\mbox{Max}\;\{|m(x)|\}$ is about $0.10$.}
\label{fig:powerLdep1}
\end{figure}

\begin{figure}
\begin{center}
\unitlength=1cm
\begin{picture}(15,3.5)
\centerline{
\epsfysize=4cm
\epsfbox{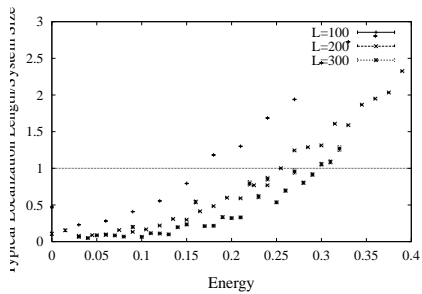}
}
\end{picture}
\end{center}
\vspace{0mm}
\caption{System size dependence of typical localization length (2):
 We set $\alpha_{pw}=0.1$. Distance between kinks is $0.5$ and
 $\mbox{Max}\;\{|m(x)|\}$ is about $0.20$.}
\label{fig:powerLdep2}
\end{figure}

In order to see whether this behaviour of the localization length
is due to the long-range correlation of $m(x)$ or the way of
introducing randomness, we performed the following calculations;
\begin{enumerate}
\item Localization length for random values of $|m(x)|$ with
exponential correlation and the fixed distances between kinks.
(We can generate an exponentially-correlated $m(x)$ 
by choosing the modified Bessel function as $c(m)$ in
 Eq.(\ref{eq:construct}) (Fig.8).)
\item Localization length for random-kink-distance $m(x)$
with fixed $|m(x)|$. 
\end{enumerate}

The results are shown in Figs.9 and 10.
In the first case, the extended states again appear at larger 
energies (see Fig.9).
We see that energy of 
the mobility edge becomes larger for larger system size.
This behaviour is the same as in the case of power-law correlated random mass. 
However, the change of the mobility edge seems very small.

On the other hand, (almost) all states are localized in the second 
case (see Fig.10).
This means that the existence of the finite mobility edge stems from
the quasi-periodicity of the random mass with {\em the distances between 
kinks fixed}.

From the above result, we conclude that the appearance of extended states
at high energy is due to the specific scheme to generate random mass and
it is independent of the correlation of $m(x)$. 
 
\begin{figure}
\begin{center}
\unitlength=1cm
\begin{picture}(15,3.5)
\centerline{
\epsfysize=4cm
\epsfbox{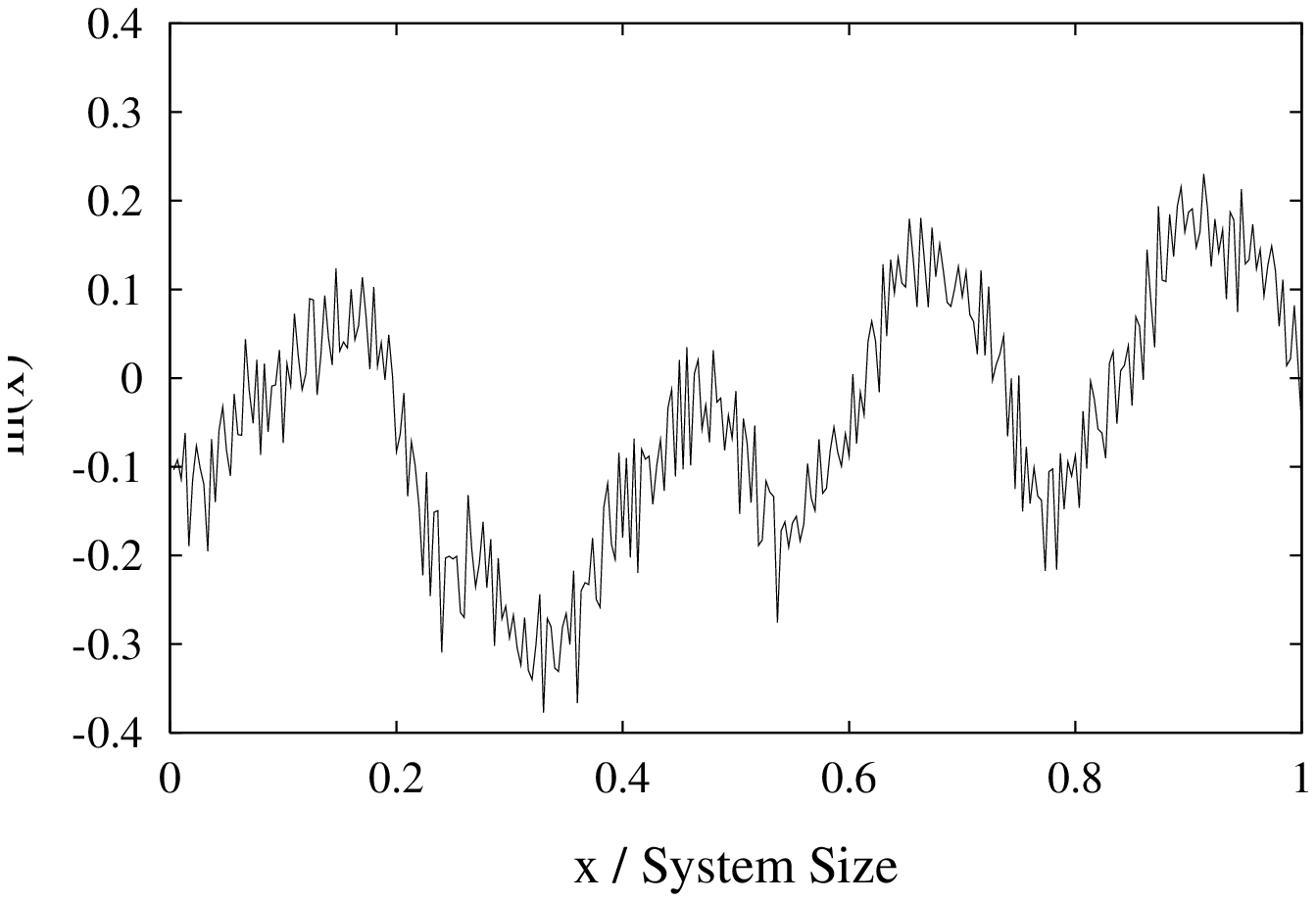}
}
\end{picture}
\begin{picture}(15,4)
\centerline{
\epsfysize=4cm
\epsfbox{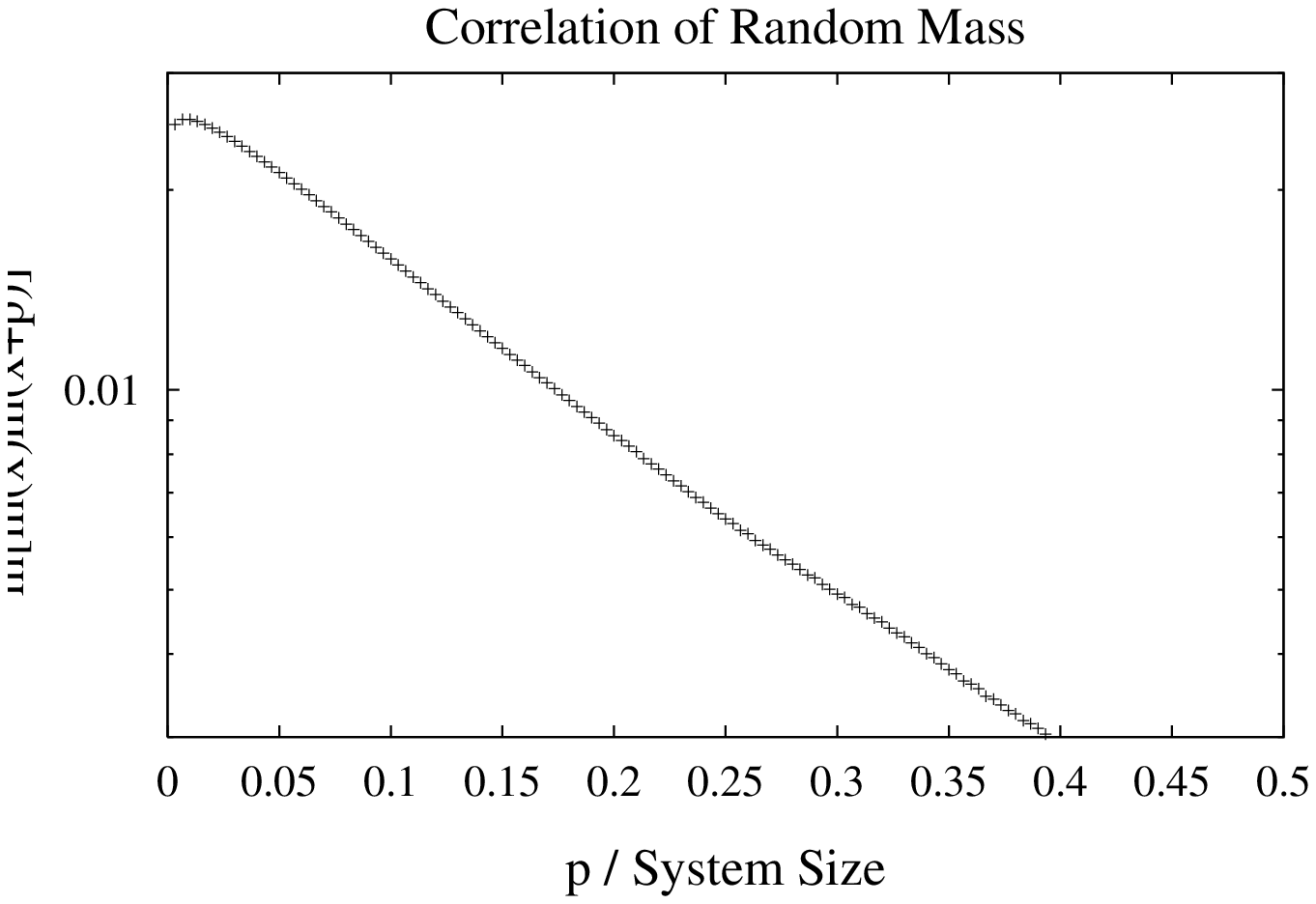}
}
\end{picture}
\end{center}
\caption{An example of $m(x)$ and the correlation:
 We found the exponentially correlated randomness
 in this example. We averaged 500 random systems for
 the calculation of correlation of random mass.}
\label{fig:corexp}
\end{figure}

\begin{figure}
\begin{center}
\unitlength=1cm
\begin{picture}(15,4)
\centerline{
\epsfysize=4.5cm
\epsfbox{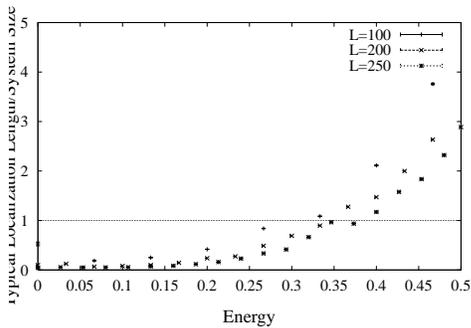}
}
\end{picture}
\end{center}
\caption{Typical localization length in the system
 with exponentially correlated 
 randomness.
 Distance between kinks is $0.333$.
 $\mbox{Max}\;\{|m(x)|\}$ is about $0.35$.}
\label{fig:exp1}
\end{figure}

\begin{figure}
\unitlength=1cm
\begin{center}
\begin{picture}(15,4)
\centerline{
\epsfysize=4.5cm
\epsfbox{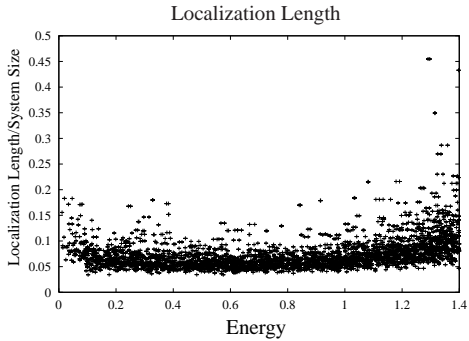}
}
\end{picture}
\end{center}
\vspace{0mm}
\caption{Localization length with exponentially correlated
randomness (with varying kink distances and constant $|m(x)|$):
 We set the correlation length $\lambda=1/2$, $|m(x)|=1$ and $L=50$.
 Localization lengths of the states do not exceed the system size even 
 in the range $E>1$.}
\label{fig:exp2}
\end{figure}

%%%%%%%%%%%%%%%%%%%%%%%%%%%%%%%%%%%%%%%%%%%%%%%%%%%%%%%%%%%%%%%%%%%%%%%
\subsection{Aperiodic random mass}

In the previous subsection, we studied the Dirac fermion with 
random mass of power-low correlation.
In order to make $m(x)$ correlated algebraicly, we fixed the distances
between kinks of the telegraphic $m(x)$ and made the magnitude of 
$m(x)$ random.
In this subsection, we shall consider the second type of random mass
in which the magnitude of $m(x)$ is fixed whereas distances between 
kinks are random variable.

Recently a closely related model, aperiodic Kronig-Penney model
in one dimension, was studied in Ref.\cite{IzKrUl}.
The Schr$\ddot{\mbox{o}}$dinger equation of the model is given as
\begin{equation}
 -(\hbar/2m)\Psi~{''}(z) + \sum^{\infty}_{n=-\infty}
 \, U \; \delta(z - z_{n}) \Psi(z) = E \; \Psi(z),
\end{equation}
where potential depth $U$ is constant.
In the periodic case, $z_{n}$'s (positions of the delta-function potential) 
are arranged periodically, that is, they satisfy 
$z_{n}-z_{n-1} = A$($=$constant). 
We then introduce randomness as the displacements of 
the delta-function potentials and
denote them $\delta_{n}$. We define the relative displacement
$\Delta_{n}$ as $\Delta_{n} = \delta_{n+1} - \delta_{n}$ and also
the correlation of $\Delta_{n}$ as
\begin{equation}
 \xi(m) = [ \Delta_{k} \Delta_{n+m} ]_{\rm ens} /
 [\Delta_{n}^{2}]_{\rm ens}.
\end{equation}
In Ref.\cite{IzKrUl}, it was shown that extended states 
(states which have vanishing Lyapunov
exponent) appear at finite energies and therefore the Anderson 
transition occurs if $\xi(m)$ decreases sufficiently slowly
according to the power law.
We apply this method of generating random distances to our model and
generate random distances between kinks by the following
scheme. 
We first generate random variables $\Delta_{n}$'s
from $\xi(m)$ by the method explained in the previous subsection (we
set $C^{'}=1$ and $k \simeq 10^{4}$ in Eq.(\ref{eq:construct}) ),
and then calculate displacements $\delta_{n}$'s by the following equation, 
\begin{equation}
 \delta_{n} = C \sum^{n-1}_{l=0} \Delta_{l},
\end{equation}
where $C$ is a parameter which controls the magnitude of the randomness.
There are three parameters in this random system, which are $A$
 ({\it mean distance} of kinks),  $C$ ({\it magnitude} of the randomness)
 and {\it decay power} of $\xi(m)$ (we denote it $\nu$) when we 
 consider the power-law correlation.

In Figs.11-13
we show $C$ and $L$(system size) dependences of the typical
localization length.
In each case the
typical localization lengths are a {\em decreasing} function of $E$ 
but becomes almost constant as $E$ gets larger
(see Figs.11-13).

In the case of {\em large} $C$, the typical localization length
for nonvanishing $E$
(normalized by the system size $L$) decreases and become smaller than unity
as we let $L$ larger.
This result indicates that almost all states are localized  
for large $C$ at the infinite system size.
We expect that this conclusion holds even for small value of $\nu$.

\begin{figure}
\unitlength=1cm
\begin{center}
\begin{picture}(15,4)
\centerline{
\epsfysize=4cm
\epsfbox{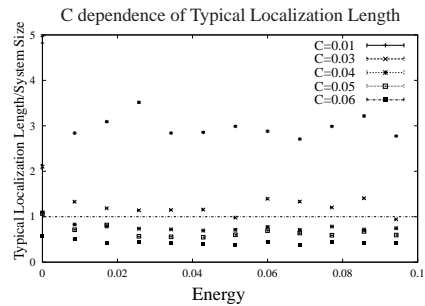}
}
\end{picture}
\end{center}
\vspace{0mm}
\caption{$C$(amplitude of randomness) dependence of typical
 localization length:
 We set $L=700$, $A=0.5$ and $\nu=-5.0$ here.
 In the case of small $C$, typical localization length 
 at any $E$ exceeds the system size.}
\label{fig:aperiod1}
\end{figure}

\begin{figure}
\unitlength=1cm
\begin{center}
\begin{picture}(15,4)
\centerline{
\epsfysize=4cm
\epsfbox{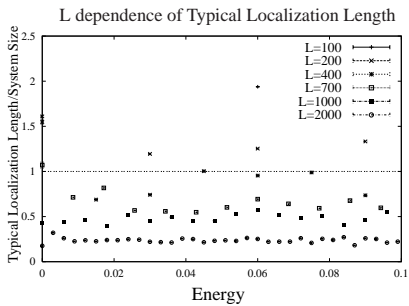}
}
\end{picture}
\end{center}
\vspace{0mm}
\caption{$L$(system size) dependence of typical localization length:
 In this figure, we show the $L$ dependence in the case of large
 $|\nu|$, which means short-range correlated random distances.
 We set $A=0.5$, $C=0.05$ and $\nu=-5.0$ here.}
\label{fig:aperiod2}
\end{figure}

\begin{figure}
\unitlength=1cm
\begin{center}
\begin{picture}(15,3.5)
\centerline{
\epsfysize=4cm
\epsfbox{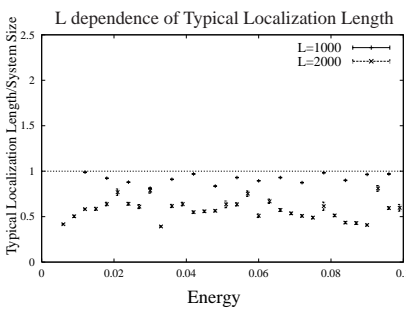}
}
\end{picture}
\end{center}
\vspace{0mm}
\caption{$L$ dependence of typical localization length:
In this figure, we show the $L$ dependence in the case of small
$|\nu|$, which means long-range correlated random distances. 
We set $A=0.5$, $C=0.04$ and $\nu=-0.5$ here.}
\label{fig:aperiod3}
\end{figure}

As the value of $C$ becomes smaller, the typical and/or mean localization
lengths exceed the system size even for very large $L$.
Dependence of $L$ is explicitly given in Fig.14.
Unfortunately at present, we cannot obtain a definite answer
to the existence of the mobility edge at finite energy.
This is in sharp contrast to the Anderson model and Kronig-Penney
model in which it is believed that a genuine Anderson transition occurs.

\begin{figure}
\begin{center}
\unitlength=1cm
\begin{picture}(15,3.5)
\centerline{
\epsfysize=4cm
\epsfbox{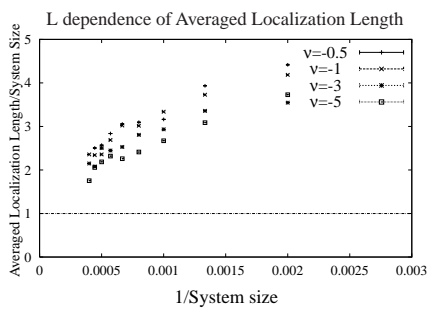}
}
\end{picture}
\end{center}
\vspace{0mm}
\caption{$L$ dependence of typical localization length:
 We set $A=0.5$ and $C=0.01$ here. We show the averaged 
 localization length of eigenstates whose energy eigenvalues
  are larger than $0.02$.}
\label{fig:aperiod4}
\end{figure}

%%%%%%%%%%%%%%%%%%%%%%%%%%%%%%%%%%%%%%%%%%%%%%%%%%%%%%%%%%%%%%%%
\section{Conclusion}

In this paper, we studied the Dirac fermion with the long-range
correlated random mass by using the TMM and IVPM.
We are especially interested in the delocalization transition or
the existence of a finite mobility edge.
Recent studies on the Anderson model and Kronig-Penney model
indicate the existence of finite mobility edges.
We studied two schemes of the random mass generation.
In the first one, magnitudes of the mass are taken to be
random variables with a long-range correlation whereas
distances between kinks are fixed.
In this case, we found the finite mobility edge.
In the second one, distances between kinks are random variables.
This case is closely related with the aperiodic Kronig-Penney model
in which the finite mobility edge is observed.
However in the present study, we cannot obtain a definite answer concerning 
the (non)existence of the finite mobility edge
because of the finiteness of the system size.
Further studies are required for that problem.

We can also calculate the multi-fractal scaling indices directly from 
wave functions of the random-mass Dirac fermion. 
Values of the indices may depend on the decay power of the 
correlation of the randomness.
Results will be reported in a future publication \cite{KI2}.

%%%%%%%%%%%%%%%%%%%%%%%%%%%%%%%%%%%%%%%%%%%%%%%%%%%%%%%%%%%%%%%%%%%%%%%%%%%%%%

%\end{thebibliography}

\end{document}